\documentclass[fleqn,10pt]{wlscirep}
\usepackage[utf8]{inputenc}
\usepackage[T1]{fontenc}
\usepackage{cuted}
\usepackage{graphicx}
\usepackage{xcolor}
\usepackage{amsmath}
\usepackage{amsfonts}
\usepackage[english]{babel}
\usepackage{amsthm}

\usepackage{esint}
\usepackage{algorithm,algorithmicx}
\usepackage{algpseudocode}
\usepackage{array}
\usepackage[caption=false,font=normalsize,labelfont=sf,textfont=sf]{subfig}
\usepackage{textcomp}
\usepackage{stfloats}
\usepackage{multirow}
\usepackage{booktabs}
\usepackage{etoolbox}
\usepackage{geometry}

\usepackage{svg}
\usepackage{lineno}

\usepackage{url}
\usepackage{verbatim}
\usepackage{cite}
\def\BibTeX{{\rm B\kern-.05em{\sc i\kern-.025em b}\kern-.08em
    T\kern-.1667em\lower.7ex\hbox{E}\kern-.125emX}}
\usepackage{subfloat}
\usepackage{tikz}
\usetikzlibrary{graphs, arrows, positioning, quotes, shapes.geometric, arrows.meta}
\title{Metasurface-Enabled Superheterodyne Transmitter for Arbitrary-Order Modulation with Spatially Isotropic Symbol Distribution}
\author[1]{Xuehui Dong}
\author[1]{Miyu Feng}
\author[1]{Chen Shao}
\author[1]{Bokai Lai}
\author[1]{Jianan Zhang}
\author[1]{Rujing Xiong}
\author[1]{Kai Wan}
\author[1,*]{Tiebin Mi}
\author[1,*]{Robert Caiming Qiu}
\affil[1]{ School of Electronic Information and Communications, Huazhong University of Science and Technology, Wuhan, 430074, China}
\affil[*]{Corresponding author: \{mitiebin, caiming\}@hust.edu.cn}

\begin{abstract}
Electromagnetically programmable information metasurfaces, as dynamically controllable 2D metamaterials, hold significant promise as low-profile hardware enabling passive wave control and signal generation for backscatter systems.
 However, current metasurface-based transmitters architecture fundamentally suffer from hardware non-modularization, forcing all transmitter functions onto nonlinear switch-based unit cells, which introduces symbol mapping inconsistency via phase coupling. Moreover, both temporal coding (limited by unit cell diodes) and space-time coding (impaired by symbol anisotropy) exhibit irreducible harmonic interference and entangled control of amplitude, phase, and beam direction.
This paper proposes a metasurface-enabled superheterodyne architecture (MSA), comprising a digital up-conversion (DUC) module performing baseband-to-intermediate frequency (IF) conversion, filtering, and digital-to-analog conversion (DAC), and a reconfigurable metasurface featuring programmable unit cells that independently control both the magnitude and phase of the reflection coefficient. Systematically, the architecture leverages a dual-stage up-conversion process—typical of superheterodyne systems—but uniquely employs the metasurface for the final RF conversion stage.
Building upon this framework, a proof-of-concept prototype featuring a 5.8 GHz magnitude-phase decoupled (MPD) metasurface (< 15° phase deviation per state) and a DAC-based DUC module is presented. Extensive validation confirms the metasurface’s capability for distortion-free mixing with arbitrary IF signals while maintaining consistent radiation patterns. The prototype successfully implements diverse QAM modulation schemes (4QAM to 256QAM) in mono-static and bi-static configurations, demonstrating symbol isotropy for spatially separated receivers and achieving a data rate of approximately 20 Mbps (at 5 MHz IF). The generated spectrograms are indistinguishable from expected patterns, indicating applicability in scenarios such as radar spoofing.

\end{abstract}
\begin{document}
%\linenumbers
\flushbottom
\maketitle
\thispagestyle{empty}

%\noindent Please note: Abbreviations should be introduced at the first mention in the main text – no abbreviations lists. Suggested structure of main text (not enforced) is provided below.

%\section*{Introduction}
The escalating complexity, cost, and physical footprint of traditional radio frequency (RF) chains, particularly in large-scale antenna arrays, present significant bottlenecks for next-generation wireless systems.
Traditionally, RF components such as mixer, power amplifier (PA), band-pass filter (BPF), and phase shifter (PS) play critical roles in signal processing.
Streamlining these bulky components by integrating RF functionality directly at the antenna element level has thus emerged as a critical research frontier.
Recent breakthroughs in new materials and device architectures offer promising solutions to simplify and integrate the RF front end. 
Approaches such as integrated circuit-based phase shifters \cite{ellinger2010integrated,uchendu2016survey,ehyaie2011novel}, tunable active materials \cite{boardman2011active,cui2019tunable}, and software-defined architectures \cite{dillinger2005software,ulversoy2010software,akeela2018software} are poised to further enhance system flexibility while reducing overall complexity.

Metasurfaces, two-dimensional arrays of reconfigurable sub-wavelength unit cells, have garnered significant interest due to their inherent structural simplicity—eschewing conventional RF components—and their remarkable ability to manipulate EM waves and encode information\cite{bose1898rotation,schelkunoff1953antennas,walser2001electromagnetic,sievenpiper2003two,dong2024wireless,zhao2020metasurface}. 
Advancements in EM metamaterials have spurred the development of metasurface-based transmitter architectures, transitioning them from wave-shifting demonstrations towards functional RF front-ends\cite{cui2014coding,zhang2018space}. Pioneering efforts utilized 1-bit phase-coding metasurfaces with PIN diodes for dynamic beam steering\cite{cui2014coding}. This limitation of low modulation orders was shattered by the advent of Space-Time-Coding (STC) metasurfaces \cite{zhang2018space}, which introduced joint spatiotemporal modulation to generate and manipulate nonlinear harmonics through Fourier-domain energy redistribution. By periodically toggling meta-atom states via FPGA-controlled sequences, these platforms achieved unprecedented spectral-spatial control—transforming passive reflectors into active signal generators. Recent breakthroughs harness STC’s inherent convolution properties for multi-user multiplexing \cite{zhang2021wireless,ke2022space}, simultaneously encoding independent data streams onto harmonically separated beams steered to distinct spatial sectors\cite{zhang2020convolution}.
Current architectures now support 256QAM millimeter-wave transmissions \cite{chen2022accurate} while eliminating conventional mixer/PA chains. As metasurface transmitters exhibit monolithic designs, they are poised to underpin next-generation joint communication-sensing networks \cite{zhang2021wireless,wan2019multichannel}, embedding electromagnetic intelligence directly into radiating surfaces.

However, this architectural simplification introduces significant coupling between the baseband signal phase and carrier phase. Consequently, the symbol-to-codebook mapping becomes inconsistent across different metasurface configurations (e.g., size, quantization levels), incident wavefront conditions, and receiver locations.
This fundamental limitation originates from the inherent lack of hardware decoupling in existing metasurface-based transmitter architectures  \cite{9048622}, specifically in what we refer to as the digital-coding metasurface based architecture (DCMA) transmitter .
In the DCMA-transmitter, the absence of conventional transmitter link modules (e.g., mixer, filter and etc.) necessitates the implementation of all transmitter functionalities through digital state switching of nonlinear components (e.g., diodes, transistor) embedded within the unit patch.
Furthermore, the DCMA-transmitter, encompassing both aforementioned schemes, suffers from significant harmonic interference and coupling effects. These phenomena result in an inherent entanglement between the amplitude, phase, and radiation direction of each harmonic component. Although recent studies\cite{luo2024fully,zhang2021wireless}, have attempted to mitigate this entanglement and exploit it for user multiplexing by serving receivers at different locations, the harmonic components cannot be completely suppressed or independently controlled due to fundamental limitations imposed by Fourier transform principles.

The DCMA-transmitter supports two fundamental coding schemes capable of concurrently generating communication signals and controlling reflected beam patterns.
The first scheme is the temporal coding which introduces periodically varying digital control signals that encode information in the time or frequency domains, generating specific time-frequency information\cite{tang2019programmable,dai2019realization,dai2019wireless,cui2019direct}.
This approach has been utilized for independent control of harmonic amplitude and phase \cite{dai2018independent}, efficient frequency conversion \cite{dai2020high}, multi-polarization conversion \cite{ke2021linear}, etc. 
Nevertheless, the modulation order of this scheme is fundamentally constrained by the number of diodes integrated within each unit cell, making high-order quantization challenging to realize in practical metasurface designs. 
To overcome this limitation, researchers have developed an alternative approach called STC\cite{zhang2018space,zhang2021wireless}. This scheme employs joint optimization of coding sequences across both spatial and temporal domains, thereby enabling higher-order modulation that surpasses the fundamental constraints imposed by the limited number of diodes per unit cell. Consequently, the primary challenge impeding the practical implementation of this scheme lies in the absence of symbol spatial isotropy. This phenomenon manifests when receivers at distinct spatial locations fail to decode identical symbols, despite employing an optimized STC matrix. While symbol anisotropy may benefit specialized applications (e.g., physically secure communications), it exhibits fundamental incompatibility with conventional public wireless systems or backscatter systems. For instance, in multipath environments, symbols propagating in different spatial directions inevitably interfere with each other, thereby degrading system utility.

\begin{figure}[h]
\centering
\includegraphics[width=0.5\linewidth]{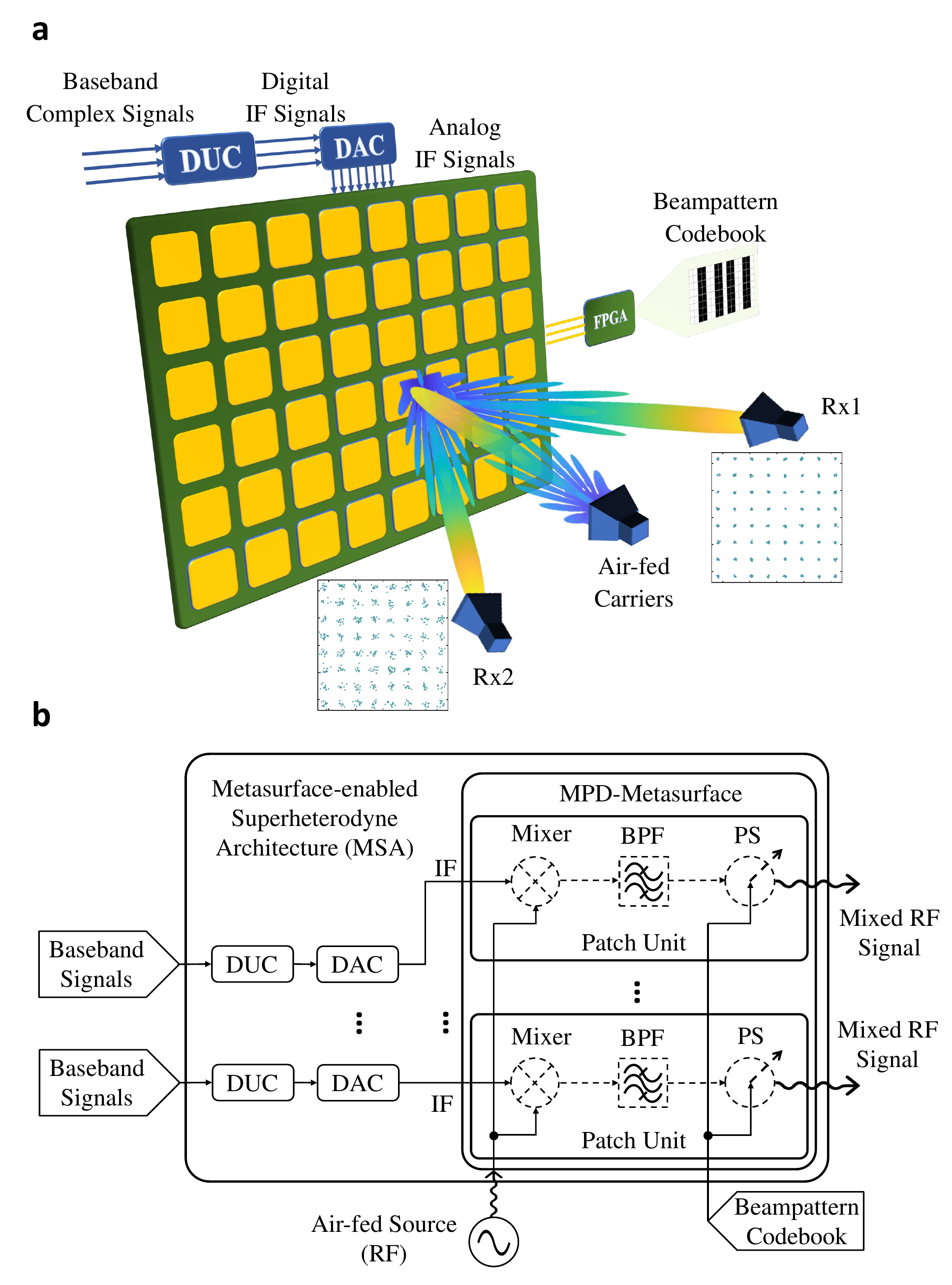}
\caption{\textbf{Conceptual illustration and Schematic diagram of MSA.} \textbf{a}, The conceptual illustration proposed MSA backscatter transmitter can independently generate and transmit arbitrary high-order complex signals while performing beamforming, ensuring spatial isotropy of the symbols. \textbf{b}, The schematic diagram of the proposed MSA backscatter transmitter. }
\label{concept}
\end{figure}

In this paper, we propose, for the first time, a novel paradigm of metasurface based transmitter architecture called as metasurface-enabled superheterodyne architecture (MSA). As shown in Fig. \ref{concept}a, the system simultaneously maintains symbol spatial isotropy while independently performing two key functions: (1) arbitrary complex signal generation and (2) beamforming operations. The MSA-transmitter consists of two primary components. The first component is a digital up-conversion (DUC) module, which performs three key functions: (1) upconverting digital baseband I/Q signals to an intermediate frequency (IF), (2) implementing IF filtering, and (3) generating analog IF signals through a digital-to-analog converter (DAC). The direct digital synthesizer in DAC provides critical advantages in modern wireless transmitter, which is responsible for the precise IF signal generation in the proposed MSA (see the architecture illustration in Fig. \ref{concept}b).
The second component comprises a reconfigurable metasurface composed of programmable unit cells, each capable of independently manipulating the magnitude and the phase of reflection coefficient. This metasurface module also performs three critical functions: (1) frequency mixing between incident RF signals and input IF signals, (2) out-of-band signal rejection through spectral filtering, and (3) programmable phase shifting of the outgoing mixed RF signal.
From a system-level perspective, this implementation follows a superheterodyne architecture through its dual-stage up-conversion process: the DUC module first translates the baseband signal to an IF, followed by a subsequent up-conversion to RF through the metasurface-based mixing operation. This cascaded frequency translation architecture provides the characteristic benefits of traditional superheterodyne systems while leveraging metasurface technology for the final up-conversion stage.

Building upon the MSA framework, we first design the magnitude-phase decoupled (MPD) metasurface operating at 5.8 GHz, whose unit maintains stable voltage-phase characteristics ($\Delta\varphi<15^{\circ}$) across two states while enabling wide dynamic range magnitude variation (i.e., $0.2$ to $1$ for state 0, and $0.1$ to $0.8$ for state 1). To experimentally validate the core functionality of the metasurface component in MSA, we conducted comprehensive arbitrary waveform generation and radiation pattern tests. These experiments demonstrate the system's capability to perform distortion-free frequency mixing between arbitrary IF signals and incident RF carriers while maintaining consistent radiation patterns. Then we develop a MSA based backscatter transmitter prototype consists of the deisgned metasurface and the DAC based DUC module, and implements diverse quadrature amplitude modulation (QAM) (e.g, 4QAM, 16QAM, 64QAM, 256QAM and etc) in both mono-static and bi-static backscatter outdoor/indoor scenarios. We further validate the symbol isotropy property, demonstrating that spatially separated receivers can successfully decode identical symbol sequences. Our prototype achieves a data rate of approximately 2$0$ Mbps when operating with an IF of $5$ MHz. We further demonstrate that the proposed MSA can generate spectrograms indistinguishable from expected patterns, enabling applications such as radar spoofing.

\section*{Metasurface-enable Superheterodyne Architecture}
\subsection*{RF Domain: Mixer and Phase Shifter}
Here we will give the theoretical basis for the idea that variations in the time and space domains of metasurfaces can be controlled independently only if certain condition is satisfied, which can instruct the design of MSA.
Firstly, we wish to express the reflected electric fields (REF) $E_{\text{r}}(\tau,\mathbf{r})$ as the multiplication of time-varying part and space-varying part, i.e. $E_{\text{r}}(\tau,\mathbf{r})=\tilde{E}_{\text{r}}(\tau)\hat{E}_{\text{r}}(\mathbf{r})$. According to array signal processing theory \cite{van2002optimum} and communication principle \cite{jacobs1965principles}, the spatial energy distribution $\hat{E}_{\text{r}}(\mathbf{r})$ constitutes the dual-domain representation of the surface phase distribution. {\color{blue} Furthermore, the time-varying component $\tilde{E}_{\text{r}}(\tau)$ carries the complete transmitted information. 
This separation indicate that the information carried in the REF is identical at any point in space domain.}
Remarkably, the temporal evolution of the phase distribution occurs at a substantially lower rate in comparison with baseband signals. We characterize the general reflection coefficient of the $n$-th unit as
\begin{equation}
    \Gamma_n(\tau, t)=\gamma_n(\tau) e^{j \theta_n(t)}
    \label{GRC}
\end{equation}
where $\tau$ and $t$ respectively denote the variables in fast-time domain and slow-time domain. 
The fast-time domain component $\gamma_n(\tau)$, termed the modulation factor, encapsulates the metasurface's rapidly tunable electromagnetic properties. These include amplitude\cite{hong2021programmable,zhang2022active}, phase\cite{yang2016programmable,kamoda201160}, amplitude and phase combined \cite{huang2021graphene}, or polarization direction\cite{huang2021polarization}, orbital angular momentum \cite{ren2019metasurface,bai2020high}, and so on.
The slow-time domain component $\theta_n(t)$, referred to as the phase factor, governs the relative phase difference between reflected and incident waves.

Conventionally, independent control of these two factors proves challenging due to the inherent coupling between the patch antenna's physical characteristics and its electromagnetic properties. However, the next section about unit design demonstrates how our proposed MPD-metasurface achieves this decoupled control, enabling command over both magnitude and phase independently. Assuming independent control of modulation and phase factors across all metasurface units, and neglecting spatial wideband effects and path loss, without loss of generality, the reflected field-incident field relationship during an instantaneous slow-time slot (see S.Eq.(5) in Supplementary Note.1 for the derivation) is
\begin{equation}
    E_r(\tau,\mathbf{k_r})=\tilde{E}_r(\tau)\bar{E}_r(\mathbf{k_r})=\underbrace{\tilde{E}_i(\tau)\sum_{n=0}^{N-1}\gamma_n(\tau)}_{\text{signal mixer}} \cdot\underbrace{\iint\text{d}\mathbf{k_i}\cdot \mathbf{w}^T \mathbf{v}(\mathbf{k_r},\mathbf{k_i})\bar{E}_i(\mathbf{k_i})}_{\text{phase shifters}}.
    \label{decouple}
\end{equation}
where $\mathbf{w}=[e^{j\theta_{0}},e^{j\theta_{1}},\dots,e^{j\theta_{N-1}}]^T$. 
Here, $\tilde{E}_r(\tau)$ represents the signal in the fast-time domain, while  $\bar{E}_r(\mathbf{k_i})$ denotes the radiation pattern in the wavenumber domain.
As demonstrated in Eq.(\ref{decouple}), the first term associated with $\tau$ functions as a signal mixer or multiplier, combining the incident signal with the superposition of rapidly varying modulation components. The second term, dependent on the wavenumber vector $\mathbf{k}$, reveals that the reflected beam patterns can be exclusively controlled through the spatial distribution of phase factors $\mathbf{w}$.
In this study, we consider an incident electromagnetic wave arriving exclusively from direction $\mathbf{k_0}$, consisting solely of carrier components (i.e. $\tilde{E}_i(\tau)=1$). Under these conditions, the REF simplifies to:
\begin{equation}
    E_r(\tau,\mathbf{k_r})= \sum_{n=0}^{N-1}\gamma_n(\tau)\cdot \mathbf{w}^T \mathbf{v}(\mathbf{k_r},\mathbf{k_0}).
    \label{single AoA}
\end{equation}

As a key theoretical basis of the proposed MSA, we highlight that the Eq.(\ref{decouple}) and Eq.(\ref{single AoA}) implies the form of signal generation and beamforming decoupling as long as the modulation factor and beam pattern factor can be independently controlled (see S.Theorem 1 in Supplementary Note.1). 
This allows for the separate manipulation of the time-domain and space-domain characteristics, thereby figures out the spatial anisotropy of symbols in STC metasurface (see the conceptual illustration in Fig. \ref{concept}a). The symbol isotropy property plays a crucial role in public wireless communication systems, as transmitted symbols propagating along different outgoing directions may interfere with each other in multipath environments—a potentially fatal phenomenon.
\subsection*{IF Domain: Superheterodyne principle}

As demonstrated in the preceding subsection, the MPD-metasurface enables up-conversion of transmitted signals to RF frequencies by ensuring that the modulation factors possess identical waveforms as signal. To achieve decoupling, the modulation factors must remain independent of the phase factors—specifically, only the magnitude of the reflection coefficient may vary. It seems like that only the amplitude modulation is available under the decoupling framework in Eq.(\ref{decouple}). To break through this limitation, we innovatively introduce the superheterodyne concept into metasurface-based transmitters, namely MSA.

Given one complex baseband symbol $x=a+jb\in\mathbb{C}$ where $a$ is the real part and the image part, the traditional zero-IF transmitter will directly up-converse it to RF by in-phase and quadrature (IQ) modulation as
\begin{equation}
    x_{\operatorname{RF}}(t)=[a\operatorname{cos}(2\pi f_{\operatorname{RF}}t)-b\operatorname{sin}(2\pi f_{\operatorname{RF}}t)]\left(u(t)-u(t-T_s)\right)
\end{equation}
where $f_{\operatorname{RF}}$ denotes the RF, $u(t)$ is the step function and $T_s$ is the symbol duration. Essentially, the $\pi/2$ phase difference between the real and imaginary parts of the carrier wave in the above equation is defined relative to the RF frequency, as is the phase factor in equation Eq.(\ref{GRC}). This direct up-conversion architecture makes it particularly challenging for metasurface-based transmitters to independently configure the symbol phase and beamforming phase, due to the inherent lack of hardware decoupling. To circumvent this limitation while satisfying the requirement for symbol isotropy, we introduce an IF concept by drawing inspiration from the architecture of superheterodyne transmitters (see the conceptual illustration in Fig. \ref{concept}b). The complex baseband symbol is first translated to IF, given by
\begin{equation}
    x_{\operatorname{IF}}(t)=[a\operatorname{cos}(2\pi f_{\operatorname{IF}}t)-b\operatorname{sin}(2\pi f_{\operatorname{IF}}t)]\left(u(t)-u(t-T_s)\right)
    \label{DUC}
\end{equation}
where $f_{\operatorname{IF}}\ll f_{\operatorname{RF}}$ thereby ensuring the symbol phase does not affect beamforming. This process, referred to as DUC in Eq. (\ref{DUC}), can be efficiently implemented and converted into a corresponding biased analog IF signal $x_{\operatorname{IF}}(t)$ using a single DAC. The $x_{\operatorname{IF}}(t)$ can be upconverted to RF by mixing with the incident carrier signal (assumed as $\operatorname{cos(2\pi f_{RF}t)}$) through the multiplier described in Eq. (\ref{decouple}), yielding:
\begin{equation}
    x_{\operatorname{RF}}(t)=x_{\operatorname{IF}}(t)\cdot\operatorname{cos(2\pi f_{RF}t)}e^{j\theta}
    \label{}
\end{equation}
where $x_{\operatorname{RF}}(t)$ denotes the reflected signal and the $e^{j\theta}$ denotes the phase factor term. By incorporating an IF, the phase of the transmitted symbols becomes distinct from that of the RF carrier, thereby achieving complete decoupling between signal generation and beamforming. This ensures that the same complex symbols are received uniformly across all spatial locations.

%\clearpage
\begin{figure}[ht]
\centering
\includegraphics[width=\linewidth]{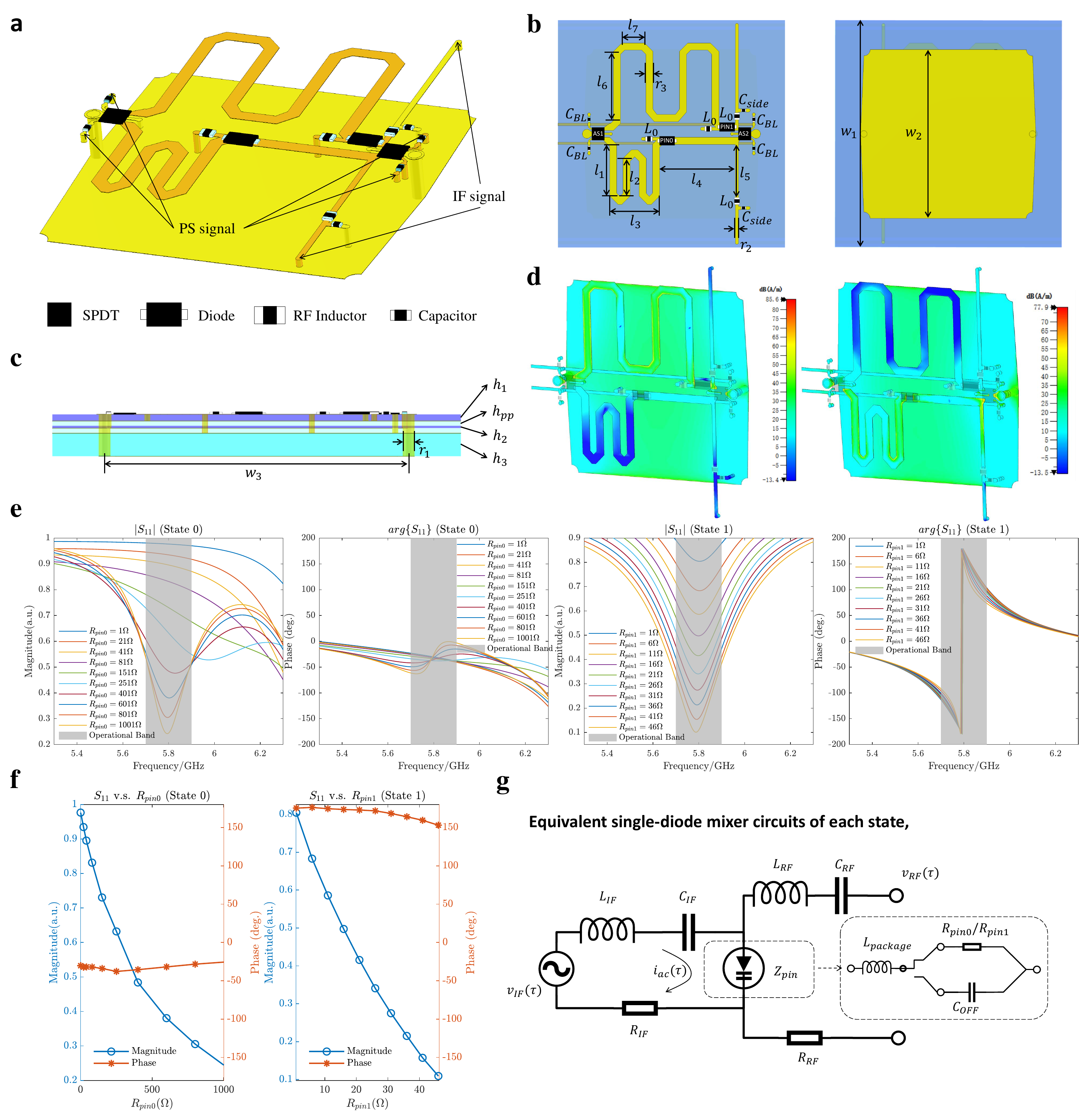}
%\caption{$\mathbf{a}$ Schematic of STD-metasurface. Analog switch joints the continuous signal and digital signal. By applying certain low-frequency (within the bandwidth, see details in Supplementary.A) continuous control signal, the outgoing EM waves will be shaped into the corresponding envelope. $\mathbf{b}$, $\mathbf{c}$ and $\mathbf{d}$ are the 1D simulation results of $S_{11}$, also known as reflection coefficients, from CST studio. $\mathbf{b}$ displays the magnitude $\vert S_{11} \vert$ of the unit under varying bias-voltages. $\mathbf{c}$ shows the phase $\angle S_{11}$ under different bias-voltages. The solid lines indicate characteristics when the PIN diode is in the conductive state, i.e., state ON. The dashed line indicates characteristics when the cross voltage is zero, i.e., OFF state. In the left subplot of $\mathbf{d}$, the relationship between phase and equivalent resistance is illustrated, while the right subplot displays the magnitude. $\mathbf{f}$ shows the comparison of 2D far-fields beampatterns of one specific configuration while the deviations of RC's phase is $0^{\circ}, 10^{\circ}, 20^{\circ}, 30^{\circ}$}.
\caption{\textbf{Hardware design of proposed narrowband magnitude-phase decoupled unit.} \textbf{a}. Geometrical configuration of the metallic unit components. \textbf{b.} Bottom view and top view of the unit. \textbf{c.} Side view of the unit. \textbf{d.} The surface current density distribution across the metallic unit components under different switching states of the SPDT RF switches. \textbf{e.} The 1D simulated reflection coefficients of unit under varying resistance. \textbf{f.} The simulated magnitudes and phases of unit at 5.8 GHz. \textbf{g.} The equivalent single-diode mixer circuits of the unit.}
\label{fig02}
\end{figure}
\clearpage

\section*{Unit Design: Reflected Mixer and Phase Shifter}
Prior research has rarely explored independent control of both magnitude and phase in metasurface unit reflection coefficients. While most existing designs operate across broadband frequencies\cite{duan2025prephase,wang2022broadband}, even narrowband implementations demonstrate different frequency selectivity that vary across phase states\cite{liao2022independent}.
To minimize out-of-band spectral pollution, the proposed design must simultaneously achieve MPD characteristics and narrowband reflection coefficient magnitude ($\vert S11\vert$) selectivity.

Building upon the MSA concept and the preceding theoretical analysis of RF-domain functionalities, we establish four design guidelines for unit reflection coefficient $S11$: (1)  Maximization of linear amplitude dynamic range for each phase state to enhance reflection modulation efficiency; (2) Phase response stability throughout the tunable amplitude range to maintain beam pattern consistency within each codebook configuration; (3) Ensure identical amplitude selectivity across different phase states; (4) Maintain consistent trends in amplitude variations.

Fig. \ref{fig02}b and Fig. \ref{fig02}c present the bottom-top view and side view of the proposed unit cell, respectively. The geometric period is designed as$w_1=18$ mm ($w_1<\lambda/2$ at $5.8$ GHz) to suppress grating lobes in the reflected electromagnetic waves.  Each unit comprises three layers, including the radiation layer, the IF signal layer, and the functional network layer. The three parts are laminated together with two
pieces of prepreg substrates. The specific dimensions of the design and the material setting can be found in Supplementary Note.2.
The unit adopts a dual-via coupling configuration, where the top-layer design incorporates tangential cuts along both sides of the rectangular patch adjacent to the vias. This innovative geometry significantly enhances the coupling efficiency of high-frequency energy into the underlying circuitry. The four corners of the rectangle are rounded to suppress undesired characteristic modes induced by sharp-edge radiation. 

In the functional network layer, by utilizing two single-pole double-throw (SPDT) RF switches, the RF signal can be controlled to pass through either the upper path or the lower path, see Fig. \ref{fig02}d for the 3D simulation results of surface current.
The phase delay can be precisely controlled through careful adjustment of the electrical length in the delay path. As illustrated in Fig. \ref{fig02}a, the upper transmission path is specifically designed with an additional $\pi$ radians (equivalent to a half-wavelength) of electrical length compared to the lower path, thereby creating the required 180° phase differential. This configuration effectively renders the SPDT RF switches functionally equivalent to a phase shifter in the system implementation. 

As shown in Fig. \ref{fig02}c, the unit cell is meticulously modeled and simulated using CST Microwave Studio 2019, with the phase and magnitude response at $5.8$ GHz. The PIN diode for State 0 (BAR64-02V) is integrated into the lower path, exhibiting an equivalent resistance range of $1$ $\Omega$ to $1001$ $\Omega$ under forward bias and a fixed parasitic inductance of $0.6$ nH. Conversely, the State 1 PIN diode (SMP1345-079LF) is integrated into the upper path, with an equivalent resistance range of $1$ $\Omega$ to $46$ $\Omega$ and an identical parasitic inductance of $0.6$ nH  (see Fig. \ref{fig02}e). The measurement results (see Fig. \ref{fig02}f) indicate that the reflection coefficient magnitudes ($\vert S_{11}\vert$) for the two operational states range from 0.8 to 0.1 and from 1 to 0.2, respectively. Furthermore, the phase responses ($\arg\{S_{11}\}$) remain stable at approximately $170^{\circ}$ and $-25^{\circ}$ for each corresponding state.

\subsection*{Equivalent single-diode mixer}
From the perspective of equivalent circuit analysis, this section elucidates the feasibility and fundamental principles of undistorted baseband signal modulation onto the incident electromagnetic wave envelope. The proposed approach employs small-signal mixers integrated into patch antenna elements utilizing PIN diodes.
Figure \ref{fig02}g illustrates the equivalent mixer circuit representation of the proposed unit design. the variables with RF subscripts in the right-side network represent the RF equivalent circuit, while the left-side loop corresponds to the equivalent circuit for IF signals. This equivalent circuit model effectively captures the key signal conversion characteristics of the implemented mixer topology. When the diode is conducting, it can be modeled as a dynamic resistor, with its small-signal relationship expressed through a Taylor expansion as follows:
\begin{equation}
    \begin{aligned}
i(\tau) & =I_s\left(e^{\alpha\left(v_{\operatorname{RF}}(\tau)+v_{\operatorname{IF}}(\tau)\right)}-1\right) \\
& \approx I_0+\frac{v_{\operatorname{RF}}(\tau)+v_{\operatorname{IF}}(\tau)}{R_d}+\frac{\left(v_{\operatorname{RF}}(\tau)+v_{\operatorname{IF}}(\tau)\right)^2}{2 R_d^{\prime}}
\end{aligned}
\label{taylor}
\end{equation}
where $R_d$ and $R_d^{'}$ respectively denotes the dynamic equivalent resistance as shown in Fig.{\ref{fig02}}f. 
By working at the proper region of the unit, the cross-product term in Eq.(\ref{taylor}) remains , as 
\begin{equation}
    i_{a c}(\tau) \approx \frac{v_{\operatorname{RF}}(\tau) v_{\operatorname{IF}}(\tau)}{R_d^{\prime}}.
    \label{mixer}
\end{equation}

Due to the nonlinearity of the diode's junction resistance, equivalently the nonlinearity of the reflection coefficient, two superimposed input signals driving the unit can result in the multiplication of these two signals. The conclusion is that by utilizing the dynamic junction resistance characteristics of the forward-biased diodes on the metasurface, IF signals can be modulated onto the outgoing EM waves. 
Moreover, to avoid signal distortion, we need to select a working range where $R_d'$ remains close to a constant, which means the first-order derivative of junction resistance $R_d^{'}$ remains relatively stable, ensuring undistorted modulation signals (see Supplementary Note.4). This information helps in setting an appropriate level range of $v_{\operatorname{IF}}(\tau)$.

\section*{Decoupled Arbitrary Waveforms Generation and Beamforming}

\begin{figure}[hb]
\centering
\includegraphics[width=0.94\linewidth]{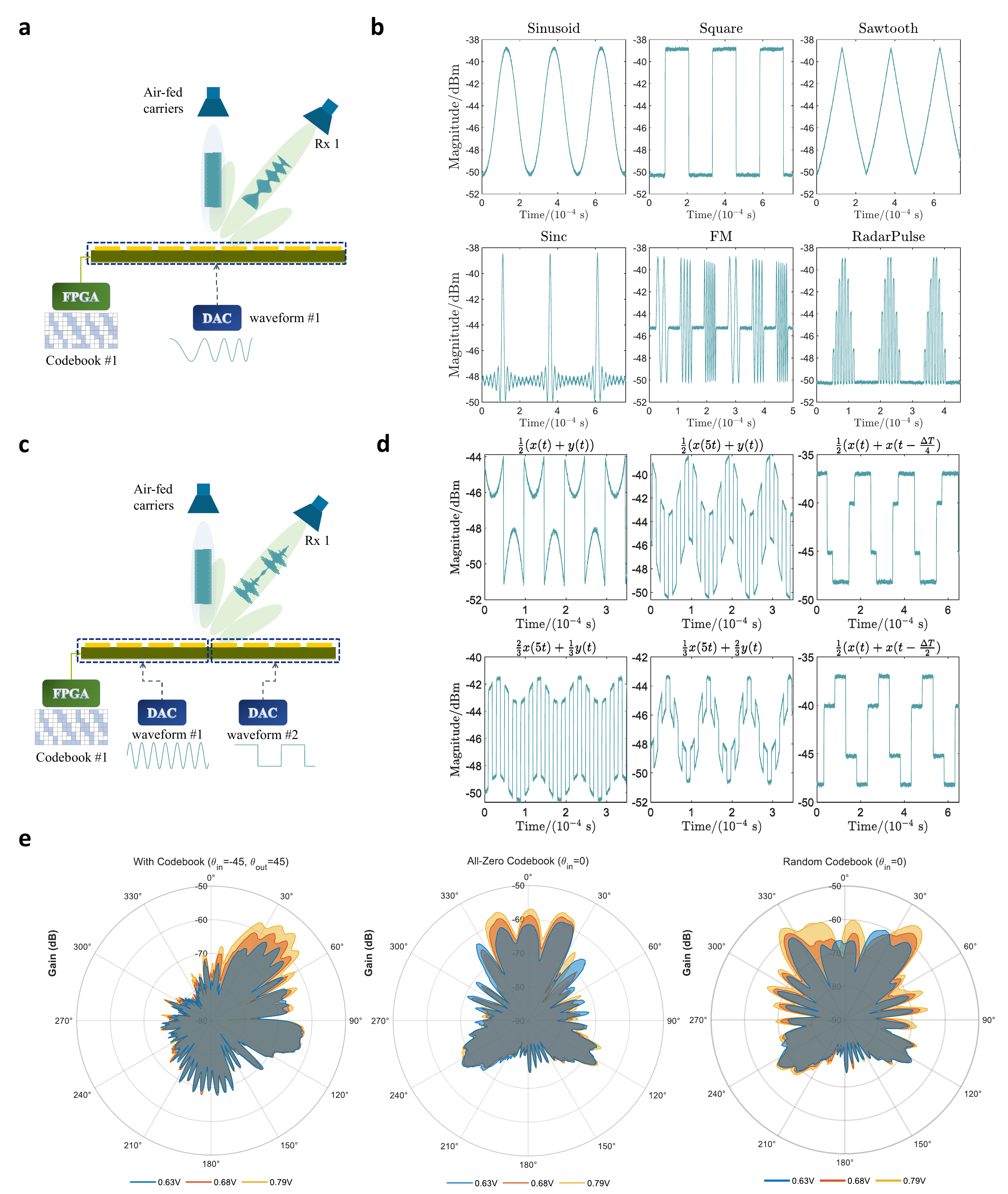}
\caption{\textbf{Experimental received waveform and measurements of radiation patterns} \textbf{a.} Illustration of arbitrary waveform generation utilizing a single DAC input. \textbf{b.} The experimental received single waveforms. \textbf{c.} Illustration of arbitrary waveform generation utilizing two DAC input. \textbf{d.} The experimental received superposed waveforms. \textbf{e.} The measured radiation patterns under different bias voltage conditions while employing distinct codebook configurations.}
\label{fig03}
\end{figure}

In this section, we present experimental results and analysis of the metasurface composed of the aforementioned MPD unit cells, demonstrating its capabilities in arbitrary waveform generation and beampattern consistency. 
For the sake of convenient, we set that the incident EM waves only comes from one direction $\mathbf{k}_0$ (i.e., $\bar{E}_i(\mathbf{k_i})=0,\forall \mathbf{k_i}\neq\mathbf{k}_0$) and has constant envelope (i.e., $\tilde{E}_i(\tau)=1$). The reflected EM field can be expressed as 
\begin{equation}
    E_r(\tau,\mathbf{k_r})=\gamma_{\operatorname{sum}}(\tau) \cdot \mathbf{w}^T \mathbf{v}(\mathbf{k_r},\mathbf{k}_0),
    \label{waveform}
\end{equation}
where $\gamma_{\operatorname{sum}}(\tau)$ represents the sum of envelopes of each reflected waves that satisfies the superposition principle. With the aid of decoupling between magnitude and phase, we can easily modulate arbitrary precise waveform onto the ambient EM waves without complicated computation and the variation of beam pattern. As shown in Fig. \ref{fig03} a, the feed horn is placed 2 m ahead of the metasurface, with the receiving horn at $45^{\circ}$ right (3 m away). Several bias analog voltage signals with classical waveforms are fed into the IF signal input of the metasurface. Meanwhile, a discrete optimization algorithm is employed for beamforming at the receiver's position. The raw received signal (without any processing) is shown in Fig. \ref{fig03}b.
AS shown in Fig. \ref{fig03}c, under the same codebook and positional configuration, we feed distinct waveforms into both sides of the metasurface and apply a series of stretching/scaling transformations. The results in Fig. \ref{fig03}d demonstrate that the superposition principle described in Eq. (\ref{waveform}) is validated.

The far-field radiation patterns of MPD-metasurface with different codebooks are illustrated in Fig. \ref{fig03}e under three different bias constant voltages at the IF signal input port. The applied bias voltages span from 0.63V to 0.79V, corresponding to magnitude ranges of 0.3 to 1 for state 0 and 0.1 to 0.7 for state 1. This voltage range was carefully selected to ensure proper device operation while maintaining the desired signal characteristics for both operational states. The specific voltage-magnitude mapping was experimentally determined to optimize the trade-off between signal amplitude and distortion. As illustrated in Fig. \ref{fig03}e, $\theta_{\operatorname{in}}$ and $\theta_{\operatorname{out}}$ respectively denotes the angle of incident and angle of outgoing,  a clear trend is observed where the beam pattern progressively contracts as the bias voltage decreases. The 0 degrees corresponds to the normal direction of the metasurface front. This experimental results validates that the signal generation will not affect the beampatterns.

\clearpage
%\subsection*{Spatial Filter}
%\begin{equation}
%\mathbf{T}_{STD}=
%    \left[\begin{array}{ccccc}
%    &\mathbf{\epsilon}^T\mathbf{v_{11}} &\mathbf{\epsilon}^T\mathbf{v_{12}} &\cdots &\mathbf{\epsilon}^T\mathbf{v_{1N}}  \\
%    &\mathbf{\epsilon}^T\mathbf{v_{21}} &\mathbf{\epsilon}^T\mathbf{v_{22}} &\cdots &\mathbf{\epsilon}^T\mathbf{v_{2N}}\\
%    &\vdots &\vdots &\ddots &\vdots\\
%    &\mathbf{\epsilon}^T\mathbf{v_{N1}} &\mathbf{\epsilon}^T\mathbf{v_{N2}} &\cdots &\mathbf{\epsilon}^T\mathbf{v_{NN}}
%    \end{array}\right]=(\mathbf{I_N}\otimes\mathbf{\epsilon}^T)\mathbf{V},
%    \label{DSTTF}
%\end{equation}

\begin{figure}[ht]
\centering
\includegraphics[width=\linewidth]{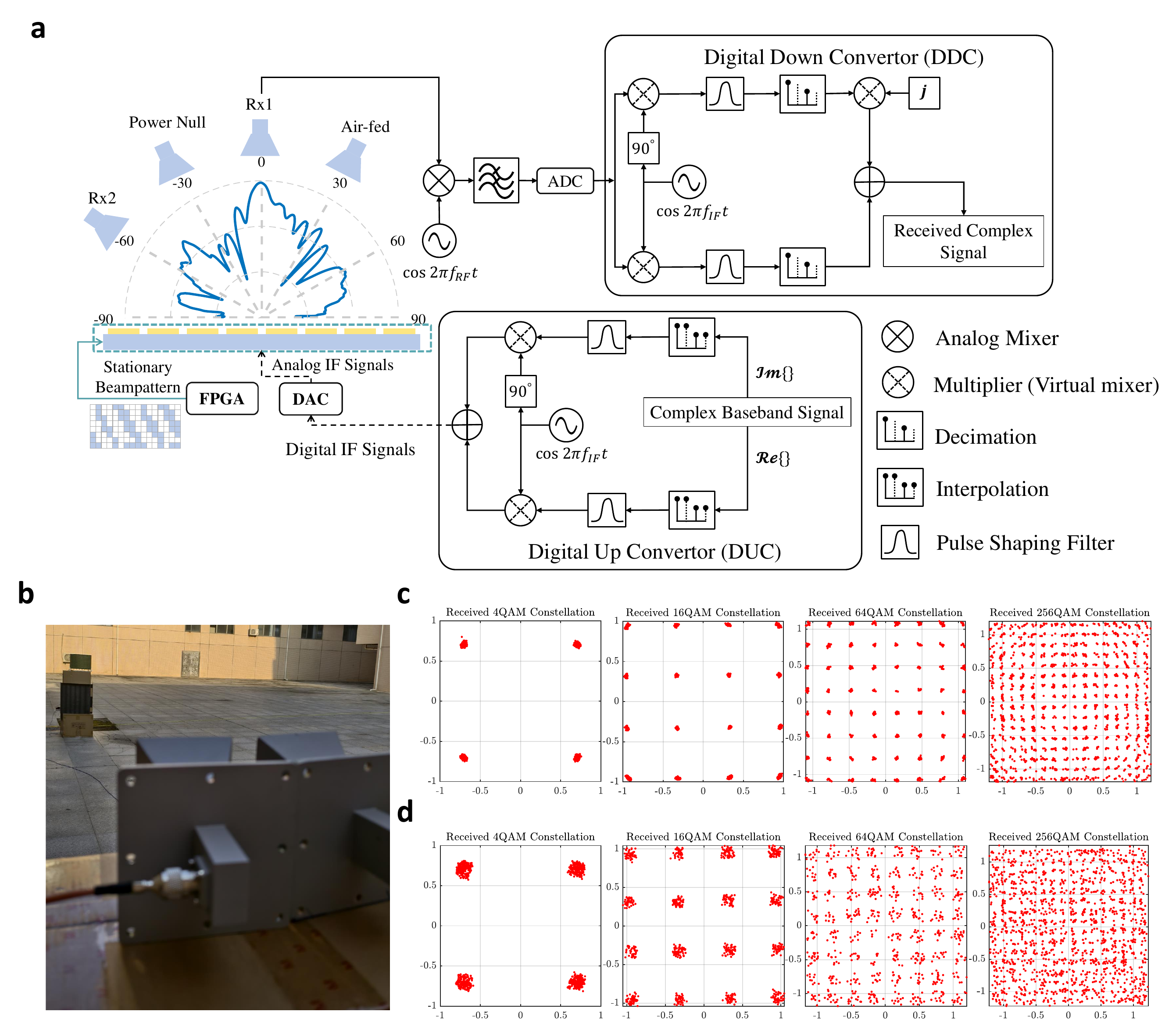}
\caption{\textbf{Schematic diagram of MSA wireless communication system and experimental received constellations} \textbf{a.} The schematic diagram of the entire system including the procedures of DUC and DDC. \textbf{b.} Experimental scenario. \textbf{c.} The received constellation diagrams at Rx1 for different modulation orders, ranging from 4QAM to 256QAM.\textbf{d.} The received constellation diagrams at Rx2 for different modulation orders, ranging from 4QAM to 256QAM.}
\label{fig04}
\end{figure}
\section*{Spatially Isotropic QAM Signals Transmission }
To validate the proposed MSA-transmitter, a realistic wireless communication system was built to perform experiments of backscatter data transmission in both indoor and outdoor environment (Fig. \ref{fig04}b). The schematic of the entire system is illustrated in Fig. \ref{fig04}a. The input bit stream is initially mapped into a complex symbol stream through constellation mapping (e.g., quadrature-amplitude modulation).The signal enters the DUC module, where the complex symbol stream is decomposed into two orthogonal components: the in-phase ($I$) branch representing the real part and the quadrature ($Q$) branch representing the imaginary part. Both streams then undergo interpolation processing based on the predetermined sampling rate and symbol rate configurations. Due to abrupt transitions between consecutive discrete symbols, high-frequency components are introduced into the signal, leading to harmful harmonics and inter-symbol interference. Therefore, the interpolated symbols are subsequently subjected to pulse shaping—here, we employ the root raised cosine filter. Subsequently, in accordance with Eq.\ref{DUC}, the $I$ signal is multiplied by the IF cosine, while the $Q$ signal is multiplied by the sine. The resulting products are then combined through subtraction ($I - Q$) to yield the digital IF signal. The digital IF signal is normalized to the appropriate voltage operating range of the metasurface, then converted into the corresponding analog IF signal via a DAC. This analog signal is subsequently fed into the designated SMA interface of the metasurface (see Fig. \ref{fig06}).

At the receiver side, the incoming signal first undergoes down-conversion through an RF mixer. It is then sampled by an ADC at the same sampling rate to obtain the received digital IF signal. Subsequently, through digital down-conversion (DDC) (the inverse process of DUC), the original baseband complex symbols can be reconstructed. As illustrated in Fig. \ref{fig04}a. The air-fed horn antenna and the two receiving horns are positioned approximately 10 m from the metasurface, with Rx1 aligned with the main lobe and Rx2 located at a sidelobe. Figs. \ref{fig04}c and \ref{fig04}d present the received constellations at Rx1 and Rx2, respectively, for various QAM modulation orders ranging from 4QAM to 256QAM. This experiment demonstrates that the MSA wireless communication system is capable of generating and transmitting signals of arbitrary high orders while maintaining spatial symbol isotropy.
Our prototype system achieves a data rate of approximately 20 Mbps, with an IF of 5 MHz and a DAC sampling rate of 20 MS/s (see Supplementary Note 7 for details).

\begin{figure}[hb]
\centering
\includegraphics[width=0.8\linewidth]{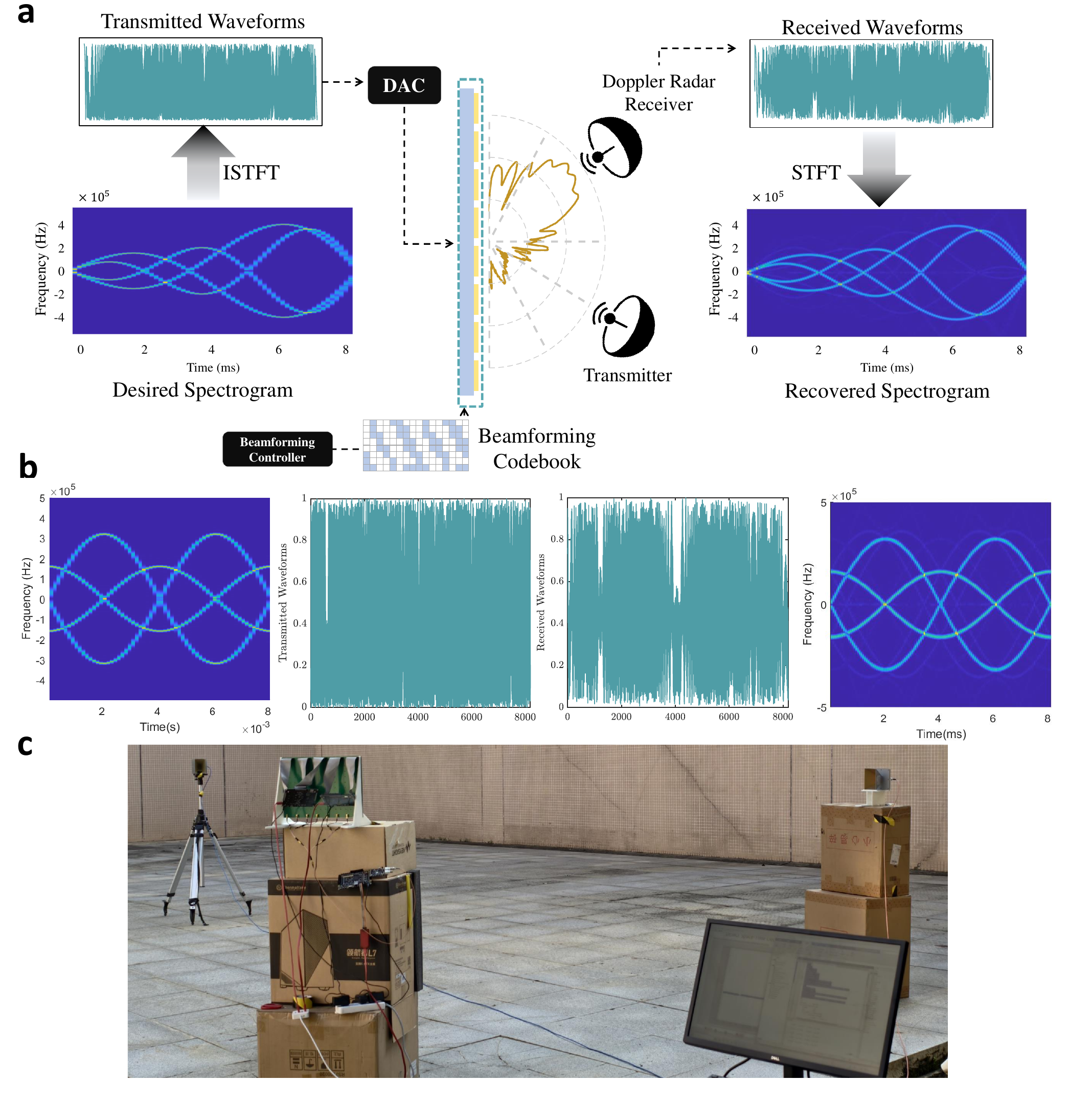}
\caption{\textbf{System diagram and Experimental spectrogram.} \textbf{a.} The system diagram of the dynamic Doppler-spoofing reflection array.
The illustration has shown that the specified spectrogram can be imitated by the metasurface. \textbf{b.} The four diagrams respectively denote the desired spectrogram of a dual rotor helicopter, transmitted waveform, received waveform and recovered spectrogram.\textbf{c.} The experimental scenario.} 
\label{fig05}
\end{figure}
\section*{Doppler Signatures Imitation}
The MSA offers significant advantages as a radar deception tag, including the ability to generate arbitrary high-precision spectra and time-frequency Doppler signatures, minimizing interception risk by eliminating leakage from sidelobe variation patterns. Additionally, it enables simultaneous generation of specific time-frequency signatures while performing beamforming to enhance or reduce the radar cross-section as required.
Metamaterials or metasurface devices are often used as radar stealth coatings or deception tags \cite{chen2014reduction,su2016ultra,kozlov2023radar}. With the advent of information metasurfaces, the flexibility and reconfigurability of metasurfaces as radar deception tags have significantly improved. Existing information metasurfaces, when used as Doppler radar deception reflection tags, face two difficult challenges: (1) Due to signal-beampattrn coupling, the spectrogram differ at various positions, making it easy to detect that the forged signal is artificial; (2) The unknown position of the radar makes it difficult to present the radar with the desired custom waveform. These two major issues greatly limit the practical use of metasurfaces as doppler-spoofing tags.

The introduction of MAS can address these two problems. As shown in Fig.\ref{fig05}a, by performing an inverse short-time Fourier transform on the desired spectrogram, the time-domain waveform can be obtained, and the corresponding waveforms can be input into the MPD-Metasurface through a DAC. Figs. \ref{fig05}a and \ref{fig05}b present spectrograms emulating helicopter and dual rotor helicopter signatures, respectively. The corresponding experimental setup is illustrated in Fig. \ref{fig05}c.

\begin{figure}[ht]
\centering
\includegraphics[width=\linewidth]{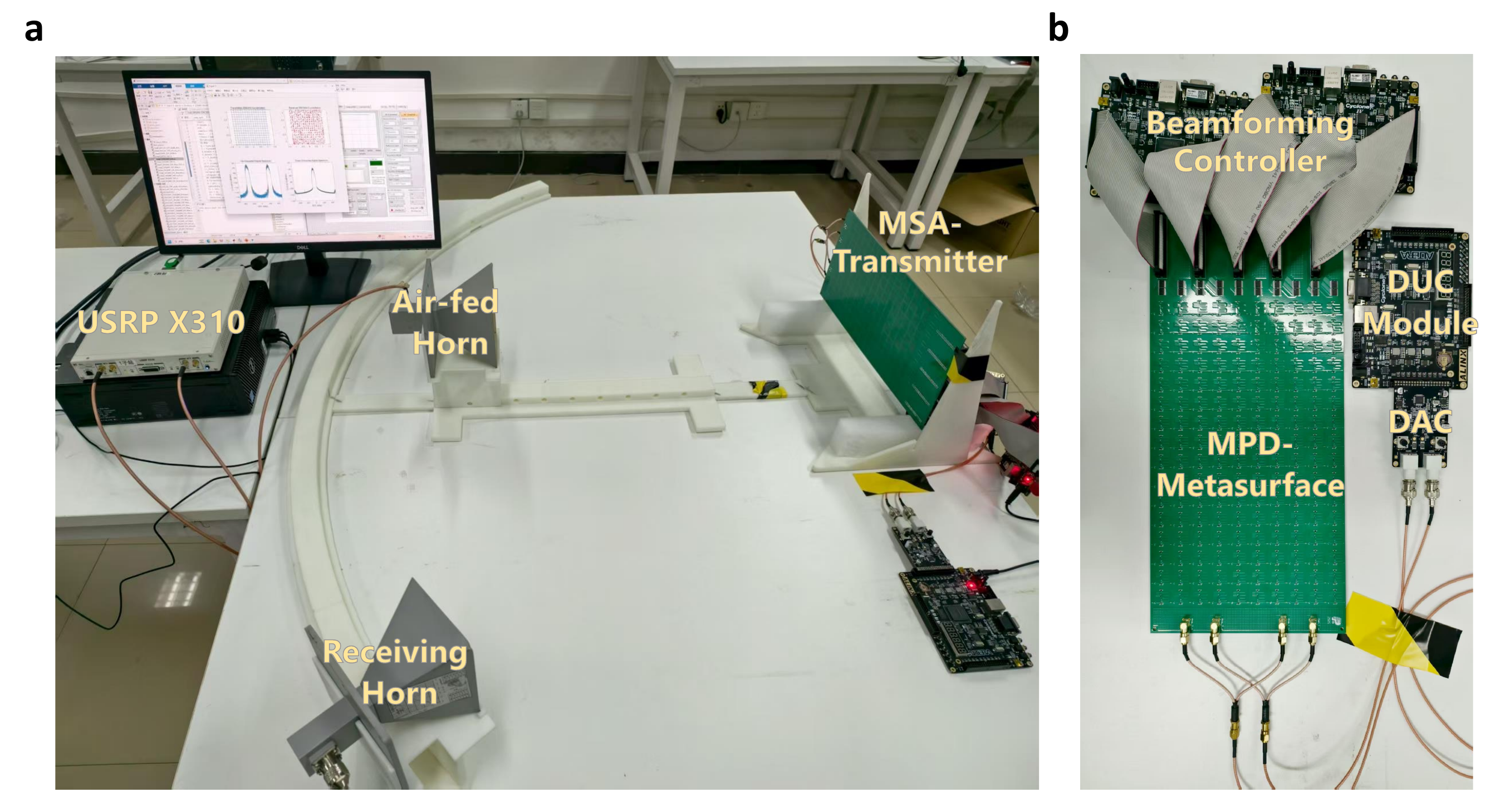}
\caption{\textbf{Experimental platform and prototype design.} \textbf{a.} The experimental platform for testing the fundamental functions of MSA-Transmitter. \textbf{b.} The composition of MSA-transmitter prototype . }
\label{fig06}
\end{figure}

%\section*{Discussion}
\section*{Discussion}
We proposed a metasurface-enabled superheterodyne architecture transmitter, comprising a DUC module and a MPD-metasurface. As a potential future wireless communication transmitter technology, the MSA framework combines the advantages of metasurfaces and direct digital signal generation. On one hand, it retains the precision and compatibility benefits of mature direct digital synthesis techniques. On the other hand, the low-profile metasurface—capable of decoupled magnitude and phase control—serves as both a reflective modulator and beamformer, replacing traditional bulky, high-cost multi-antenna RF front-ends. Furthermore, our MSA prototype experimentally demonstrates independent arbitrary-order complex signal generation and beamforming, spatial isotropy of symbols, and pulse shaping functionalities, etc. This architecture has demonstrated promising potential for future applications in wireless transmitter, backscatter communications, and sensing technologies.

Although the proposed MSA transmitter demonstrates promising capabilities, several limitations remain to be addressed. First, the maximum achievable data rate is fundamentally limited by the metasurface architecture. Enhancing this performance metric will require improvements in both high-speed control electronics and the switching characteristics of the individual meta-atoms. Second, increasing the phase quantization bit depth of the unit cells while maintaining phase stability across the dynamic magnitude variation range. In addition to hardware enhancements, we recommend integrating adaptive impedance matching networks with real-time digital predistortion algorithms to improve phase stability under dynamic magnitude variations. Further, the modulation efficiency of the metasurface can be enhanced by transitioning from double-sideband to single-sideband modulation through optimized circuit-level and system-level designs. This improvement would significantly boost the reflective modulator's performance while maintaining spectral efficiency.

\section*{Methods}

\textbf{Experimental platform and MSA-transmitter prototype.}
The MSA transmitter used in this work operates around the central frequency of $f_c =5.8\text{GHz}$ and contains $16\times 10$ magnitude-phase decoupled units. It is designed by the commercial software CST Microwave Studio and fabricated with printed circuit board technology. 
All the data in the previous sections were collected from the experimental platform (see Fig.\ref{fig06}(a)). We built the system with aid of SDR device (NI USRP-X310) and the post-processing computer.
The prototype (see Fig.\ref{fig06}(b)) consists of the MPD-Metasurface, the beamforming controller (two AX515 FPGAs) and  the DUC module (one AN9767 DAC and one AX515 FPGA). For the sake of convenience, we used several splitters for ensuring the input voltage signal is identical. The integration of two diode types within each unit cell results in partially overlapping linear operating regions. To maximize system efficiency, we implemented dedicated IF control interfaces for each diode type.

\bibliography{ref}

\section*{Acknowledgments}

\section*{Author contributions statement}
R.C.Q., X.D., T.M., K.W. conceived the idea; X.D., J.Z, R.X conducted the theoretical analysis; X.D., M.F., C.S. conducted the experiments and data processing; X.D., M.F., C.S., B.L. assisted in building up the system; X.D. wrote the manuscript. R.C.Q., T.M., K.W. provided suggestions and helped to organize and revise the draft. All members contributed to the discussion of the results and proofreading of the manuscript.
\section*{Competing interests}
The authors declare no competing interests.
\section*{Additional information}
Supplementary document and Four videos of demos have been submitted as attachments.

\end{document}